\documentclass[pra,aps,prl,twocolumn,superscriptaddress]{revtex4-1}
\usepackage{braket}
\usepackage{amssymb}
\usepackage{amsmath}
\usepackage{epsfig}
\usepackage{color}
\usepackage{graphics, graphicx}
\usepackage{bbold}
\usepackage{psfrag}
\usepackage{mathcomp}
\usepackage{subfigure}
\usepackage{verbatim}
\usepackage[colorlinks, citecolor=blue]{hyperref}
\usepackage[normalem]{ulem}
\usepackage{bm}
\def\cp#1{\mathbf{#1}}
\begin{document}

\title{Wetting of quantum fluids: a route to free-standing shell-shaped quantum droplets}
\author{Francesco Ancilotto}
\affiliation{Dipartimento di Fisica e Astronomia ``Galileo Galilei''
and CNISM, Universit\`a di Padova, via Marzolo 8, 35122 Padova, Italy}

\date{\today} 

\begin{abstract}
We investigate wetting phenomena between self-bound quantum 
fluids in a three-component Bose mixture of $^{23}$Na, $^{39}$K, and $^{41}$K atoms. 
Within a density-functional approach including mean-field interactions 
and Lee-Huang-Yang quantum-fluctuation corrections, we consider 
two binary quantum liquids, formed by components $(1,2)$ and $(2,3)$, and 
study the adsorption of the softer $(1,2)$ liquid on a stiffer $(2,3)$ substrate. 
By tuning the interspecies scattering length $a_{12}$, we show that the 
surface tension of the $(1,2)$ liquid can be strongly varied, 
driving a transition from partial wetting to complete wetting of the $(2,3)$ phase. 
The contact angle extracted from cylindrical-cap geometries decreases 
continuously with increasing $a_{12}$ 
and vanishes near a critical value $a_{12}^{c}= -42\,a_0$. 
In the complete-wetting regime, a finite amount of $(1,2)$ liquid 
wraps around a spherical $(2,3)$ droplet, 
producing a self-bound core-shell droplet without external confinement, 
whose component-1 density has a shell-like, hollow projection.
We further show that 
such shell-shaped quantum droplets can sustain quantized 
vortical excitations. These results identify wetting as a 
route to engineering free-standing shell-shaped quantum liquids and 
suggest new possibilities for studying capillarity, topology, 
and superfluidity in multicomponent quantum droplets.
\end{abstract}

\maketitle

\section{Introduction}

Quantum droplets, first predicted theoretically by Petrov in 2015 \cite{petrov_15} 
and subsequently observed experimentally 
in dipolar gases \cite{kadau,ferrier,Schmitt2016,Ferlaino2016}
and in homonuclear \cite{Cabrera2018,Semeghini2018} and heteronuclear 
\cite{Derrico2019,Guo2021} binary mixtures of bosonic atoms, have progressively gained 
attention due to their unique properties \cite{review}.

These self-bound states are stabilized by quantum fluctuations, 
typically described by the Lee-Huang-Yang (LHY) correction \cite{lhy}, 
which counteracts the mean-field attraction that would otherwise 
lead to collapse. 
They have superfluid properties, and are characterized 
by ultralow equilibrium densities and finite (albeit extremely small) 
surface tension, and therefore 
exhibit liquid-like properties while maintaining quantum coherence \cite{anci_mult,anci23}. 

The physics of quantum droplets has been mainly
investigated in single- and two- component dipolar gases and two-component Bose mixtures \cite{review}.
Recently, 
the possibility of realizing self-bound states in three-component Bose systems 
has been proposed.
Motivated by recent theoretical work on the formation of 
“Borromean” atomic clusters in three-component ultracold bosons \cite{borromean}, 
a new type of shell-shaped
Bose-Einstein condensate with a self-bound character has been proposed \cite{ma}, made of a three-component 
$^{23}$Na $^{39}$K $^{41}$K 
Bose mixture (species (1,2,3) in the following),
where the mixtures (1,2) and (2,3) both form self-bound droplets. In the proposed
system an outer shell of liquid (1,2) envelops a spherical 
core made of the (2,3) liquid, and this structure was claimed to be stable 
without the need of any trapping potential.
As shown in Ref. \cite{comment,anci25}, however, it turns out that 
the structures described in Ref. \cite{ma} are not actually the ground-state
solutions to the system but rather metastable states corresponding to 
local energy minima. The lowest-energy states for the configurations 
studied in Ref. \cite{ma} are instead "quantum dimer" configurations where
two droplets (made of the (1,2) fluid and (2,3) fluid, respectively)
are bound together by the shared component 2 \cite{comment}.

\begin{figure}[t]
\includegraphics[width=8cm]{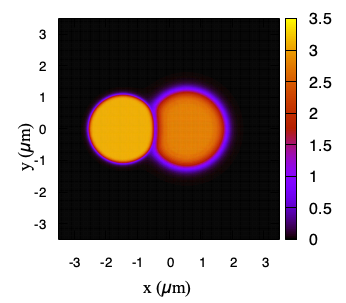}
\caption{Total density map corresponding to the dimer
structure. Here $a_{12}=-50\,a_0$.
The density values are in units of $10^4 \mu m ^{-3}$.
The droplet (2,3) is on the left, the droplet (1,2) on the right.
} 
\label{fig1}
\end{figure}

One such dimer structure \cite{anci25} is shown in Fig.\ref{fig1}, where the total density of the 
system is shown, in the plane passing through the centers of the droplets.

The (1,2) droplet on the right in Fig.\ref{fig1} is made of $N_1=3.5\times 10^4$ and
$N_2=5.9\times 10^4$ atoms, 
while the (2,3) droplet is made of
$N_2=5.4\times 10^4$ and $N_3=2.5\times 10^4$ atoms.
In the interior of each droplet, the equilibrium density ratio
$\rho _j/\rho_i $ is locked at the value $\sqrt{g_{ii}/g_{jj}}$,
as expected from the theory of binary Bose mixtures \cite{petrov_15,ma} (see also 
Section III in the following).
The interaction strengths $g_{ij}$ are those reported in Section II.

It appears that the fluid (1,2) (droplet
on the right), rather than uniformly ”wetting” the inner
spherical core of liquid (2,3) (which would result in the formation of
a shell-shaped (1,2) droplet), 
prefers instead a non-wetting configuration where it minimizes 
the contact with the (2,3) fluid surface (droplet on the left). 
Notice that the stiffer and denser droplet made of (2,3) species 
maintains its spherical shape, at variance with the softer droplet made 
of (1,2) species, whose contact surface is deformed to adapt to the (2,3) surface.

Building on such apparent "non-wetting" behavior of the (1,2) fluid adsorbed on a
rather undisturbed (2,3) core,
we explore here the possibility of changing, via fine tuning of the $a_{12}$ scattering
length, the interaction between the two subsystems: values of $a_{12}$ closer to the value expected for
the mean-field collapse of the (1,2) system would result in less dense and softer 
(i.e. with a lower surface tension) (1,2) fluid, making it possible to reach a condition where 
a wetting transition of the fluid (1,2) on a (2,3) effective substrate occurs.

We show below that this scenario can indeed be realized, and identify a critical value $a_{12}^c$ 
for such a transition to occur. We will characterize in detail this wetting transition and find, as a consequence 
of it, that a shell-shaped self-bound (1,2) droplet (i.e. stable in vacuum, without 
resorting to any confining potential) can develop, wrapping a spherical
core made of the stiffer (2,3) liquid.


Experimentally, shell-shaped Bose-Einstein condensates have been 
realized in a microgravity environment and using suitable “bubble” traps, thus
avoiding the gravitational sag that prevents a closed shell 
geometry on Earth \cite{carollo1}. A complementary realization uses
dual-species condensates in the immiscible
regime, trapped by a spherical harmonic potential: 
in a repulsive Na-Rb binary mixture,
the sodium component forms a closed outer shell while 
the rubidium component occupies the inner core \cite{jia}.
Collective excitations 
have been used to probe the hollowing transition in such 
dual-species shell condensates \cite{huang1}.

In general, 
complex microgravity-dependent magnetic traps are required 
because gravity and trapping imperfections distort
the equilibrium geometry; self-bound droplets, instead, 
fall freely as a whole, so
different components are less prone to relative gravitational sag, and 
the droplets maintain the hollow shell structure and shape in free space,
allowing researchers to study them without distortions.



The trapped-shell systems described above differ 
from the self-bound wetting-induced droplet described in the following: 
in the former, the shell structure is imposed by an external 
potential or by harmonic confinement plus interspecies 
repulsion \cite{sun,rhyno}, whereas in the present work the 
shell structure arises naturally from interfacial wetting between two 
self-bound quantum liquids and persists without an external trap.

Producing stable shell-shaped droplets would create a new kind of 
free-standing hollow quantum liquid, where curved geometry, topology, surface tension, 
and quantum-fluctuation stabilization all become experimentally accessible in one system.
These droplets enable for instance the exploration of quantum effects in non-trivial 
topologies, such as curved-space physics \cite{salasnich,tononi}. 
One example is provided in the following, where vortex structures 
in a shell-shaped, self-bound droplet will be studied.

The remainder of this paper is organized as follows: In Section II, 
the theoretical model and the numerical methods used for the
calculations are presented. 
Section III describes the obtained results.
Finally, Section IV contains a summary and outlook for future research directions.

\section{Method}

The system under study here is the same as in Ref. \cite{ma,anci25}, 
i.e. a three-component $^{23}$Na $^{39}$K $^{41}$K Bose mixture,
at zero temperature and in the absence of three-body recombination effects.
An inhomogeneous mixture made of the above species is described within 
the density functional theory (DFT) approach in the MF+LHY framework, where
the total energy functional is given by 

\begin{widetext}
\begin{equation}
E = \sum_{i=1}^{3}\int d\bm{r} \, \frac{\hbar ^2}{2m_i}|\nabla \psi_{i}(\bm{r})|^2  
+\frac{1}{2}\sum_{i,j=1}^{3}g_{ij}\int d\bm{r} \, \rho_{i}(\bm{r})\rho_{j}(\bm{r}) 
+\int d\bm{r} \, {\cal E} _{LHY}(\rho_1(\bm{r}),\rho_2(\bm{r}),\rho_3(\bm{r})) 
\label{eq:energy2c}
\end{equation}
\end{widetext}

Here $\rho_i(\bm{r})=|\psi _i(\bm{r})|^2$ 
represent the number density of each component
($i=1$ for $^{23}$Na, $i=2$ for $^{39}$K and $i=3$ for $^{41}$K). 
The coupling constants between species $i$ and $j$ are 
$g_{ij}=2\pi \hbar ^2 a_{ij}/m_{ij}$,
with scattering length $a_{ij}$ and reduced mass $m_{ij}=m_im_j/(m_i+m_j)$. 
The number densities $\rho_i$ are normalized such
that $\int _V \rho_i(\bm{r})\,{\rm d}\bm{r} =N_i$ $(i=1,2,3)$, where $N_i$ are the 
total number of atoms in the $i$-th component.

Components 1 and 3 interact via strong repulsive potential and are
therefore immiscible, while the binary mixtures (1,2) and (2,3) 
separately form self-bound binary droplets. As shown in Ref. \cite{ma}, this 
is achieved with 
$(a_{11},a_{22},a_{33},a_{23},a_{13})=(52, 30, 63, -200, 213)a_0$ ($a_0$ is the Bohr radius),
while $a_{12}$ (which is tunable via a Feshbach resonance) will be considered in the following
as variable at will, and will determine the binding properties between the
two fluids (1,2) and (2,3).
We will consider here the value $a_{23}=-200\,a_0$.
This is particularly appropriate in the present context because such a large value,
as shown in the following,
provides a stiff (2,3) system, with a relatively high surface tension. As a result,
a "substrate" made by the fluid (2,3) will prove to be rather undeformable upon 
adsorption of (1,2) fluid samples.

The term accounting for quantum fluctuations is 
\begin{equation}
{\cal E} _{LHY}=\int \frac{d^3{\bf k}}{2(2\pi)^3}\left[\sum_i(E_{i{\bf k}}-
\epsilon_{i{ k}}-g_{ii}\rho _i)+\sum_{ij}\frac{2m_{ij}g_{ij}^2\rho _i \rho _j}{\hbar ^2{\bf k}^2}\right] \label{E_qf}
\end{equation}  
with $\epsilon_{i{ k}}=\hbar ^2 k^2/2m_{i}$, and $E_{i{\bf k}}$ is the $i$-th Bogoliubov excitation energy
\cite{borromean}. 
The three dispersion relations $E_{i{\cp k}} \ (i=1,2,3)$ 
are the roots of the following equation \cite{borromean}
\begin{equation}
x^3+bx^2+cx+d=0
\end{equation}
where
\begin{widetext}
\begin{eqnarray}
b&=&-\sum_i \omega_{i}^2,\\
c&=&\sum_{i<j}\left((\omega_{i}\omega_{j})^2-4g_{ij}^2\rho _i \rho _j\epsilon_{i{ k}}\epsilon_{j{ k}}\right),\\
d&=&-(\omega_{1}\omega_{2}\omega_{3})^2-16\epsilon_{1{ k}}\epsilon_{2{ k}}\epsilon_{3{ k}}
g_{12}g_{23}g_{13}\rho_1 \rho_2 \rho _3+\sum_{i<j, l\neq(i,j)}4\epsilon_{i{ k}}
\epsilon_{j{ k}}\rho _i \rho _j g_{ij}^2\omega_{l}^2.
\end{eqnarray}
\end{widetext}
Here $\hbar \omega_{i}=\sqrt{\epsilon_{i{ k}}^2+2g_{ii}\rho _i\epsilon_{i{ k}}}$ ($i=1,2,3$) 
are the Bogoliubov spectra for the individual components. 

Minimization of the action associated to Eq.\eqref{eq:energy2c} leads
to the following Euler-Lagrange equations ($i,j=1,2,3$):
\begin{equation}
\begin{split}
i\hbar \frac{\partial \psi_i (\vec r, t)}{\partial t}=
\left(-\frac{\hbar ^2\nabla^2}{2m_i}+\sum_{j}g_{ij}\rho _j+\frac{\partial 
{\cal E}_{LHY}}{\partial \rho_i}\right)\psi_i (\vec r, t)
\end{split} \label{GP}
\end{equation}

The numerical solutions of Eqs.\eqref{GP} provide
the real-time evolution of the system in three-dimensional space.
When stationary states are sought the left hand side of
Eq.\eqref{GP} is replaced by $\mu _i\psi_i (\vec r)$,
where $\mu _i$ is the chemical potential of the $i$-th species. 
The evolution in imaginary time (via, e.g., steepest descent algorithm)
allows to obtain stationary state solutions starting 
from suitable initial wavefunctions. The chemical potentials 
$\mu _i$ are determined iteratively so
that the target values of $N_i$ are achieved. 

The wave functions $\{\psi_i \}$ are mapped on an
equally spaced 3D Cartesian grid and the Laplacian operator in Eq.\eqref{GP}
is represented by a 13-point finite-difference stencil \cite{Ral60}.
The size of the computational cell 
is large enough that the densities at the boundaries are negligible.
Periodic boundary conditions (PBC) are imposed along the three spatial directions.
The effect of possible spurious interactions between periodically repeated images 
will be discussed in Section III.B.

\section{Wetting of ($^{23}$Na,$^{39}$K) on ($^{39}$K,$^{41}$K) }

\subsection{Wetting geometry and physical setting}

In the following,
the numbers of atoms are chosen to approximately satisfy
the equilibrium ratio \cite{petrov_15} in the interior of the (1,2) and (2,3) self-bound systems
(made respectively of ($N_1,N_2^{(1)}$) and ($N_2^{(2)},N_3$) atoms), i.e. 
$N_1/N_2^{(1)}\sim \sqrt{g_{22}/g_{11}}\sim 0.583$, and 
$N_2^{(2)}/N_3\sim \sqrt{g_{33}/g_{22}}\sim 0.707$.

Notice that, for the choice of the parameters made here,
the mean-field collapse for the uniform mixture is expected to occur 
when $g_{12}=-\sqrt{g_{11}g_{22}}$, corresponding to $a_{12} \sim -39\,a_0$. More negative values of $a_{12}$
will result in the formation of a quantum liquid system.

A wetting transition \cite{bonn09} is a surface phase transition 
where a liquid changes from partial wetting (where its equilibrium shape is a
hemispherical cap adsorbed on a planar solid substrate) 
to complete wetting (where it spreads to fully cover
the surface with a thin liquid film),
usually triggered by temperature changes, causing the contact angle 
$\theta $ formed by the droplet with the substrate to drop to zero. 
The contact angle $\theta $ is the angle formed 
at the junction of liquid, solid, and gas, 
serving as a key quantitative measure of this wettability.
One usually distinguishes three regimes: $\theta =0$ (complete wetting),
$0<\theta <90^\circ $ (partial wetting), and $\theta >90^\circ $ (drying).

In the partial-wetting regime, the
contact angle is determined by balancing the forces acting 
along the contact line and depends on the interfacial
tensions $\sigma _{ij}$ between each pair of 
coexisting phases through Young's equation

\begin{equation}
 cos(\theta )=\frac {\sigma _{SV}-\sigma _{SL}}{\sigma _{LV}}
\label{young}
\end{equation}
The subscripts $L$, $V$ and $S$ identify the liquid, vapor and solid, respectively.

Young’s law is not fundamentally a “classical-liquid law.” 
It is an equilibrium capillarity condition, and as such has been used to characterize 
adsorption properties of the prototypical quantum liquid, i.e. $^4$He, on heavy alkali metallic surfaces
\cite{wet-he_exp, wet-he_th}.


Wetting phenomena have also been studied directly in Bose-condensed mixtures. 
In particular, Gross-Pitaevskii-based analyses have shown that binary 
condensates may display rich wetting phase diagrams, 
including first-order wetting, critical wetting, and prewetting transitions \cite{ind04,sch15}. 
More recently, the interface-potential approach has been used to 
characterize the effective interaction between interfaces and the 
associated line tension in Bose-Einstein-condensate mixtures 
near a hard wall \cite{sch22,song}. These works provide the natural capillarity 
framework for the present study, although here the role of the 
substrate is played not by an external wall but by another self-bound quantum liquid.

In the following, the softer (1,2) quantum liquid plays the role of the adsorbed fluid
(subscript $L$ in Eq.\eqref{young}), 
while the much stiffer (2,3) quantum liquid acts as an effective substrate
(subscript $S$ in Eq.\eqref{young}). 
Since the calculations are done at T=0,
the vapor $V$ phase will be vacuum.
This terminology is meant in the capillary sense: the (2,3) phase is 
not a solid, but its surface is only weakly deformed 
by the adsorption of the softer phase, as verified below.

\subsection{Surface tension of the adsorbed $(1,2)$ liquid}

In order to compute the surface tension $\sigma _{12}$ (corresponding 
in our model to $\sigma _{LV}$) we consider
a planar interface between the (1,2) liquid and vacuum,
as a function of the interspecies interaction strength represented by $a_{12}$.
In order to compute it, a slab geometry is used here, where 
a rectangular simulation box contains a liquid slab in the middle, of surface area $L_x L_y$
with vacuum regions on both sides along the $z$ direction.
Periodic boundary conditions are applied in all directions, so the slab geometry results
in two planar liquid–vapor interfaces.

The surface tension is calculated as

\begin{equation}
\sigma _{12}=\frac {1}{2}(\frac {E-N\epsilon _{b}} {L_xL_y})
\end{equation}
where $E$ is the calculated total energy of the slab, $N=N_1+N_2$,
$L_xL_y$ is the surface area of each of the two planar surfaces delimiting the slab,
and $\epsilon _{b}$ is the energy of a uniform phase 
with densities $\rho _1$ and $\rho _2$:

\begin{equation}
\begin{split}
\epsilon _{b} =&[\frac {1}{2} g_1\rho_1^2 +\frac {1}{2} g_2\rho_2^2 
+ g_{12}\rho_1\rho_2 \\
&+ {\cal E} _{LHY}(\rho_{1},\rho_{2})]/(\rho_1+\rho_2) 
\end{split}
\end{equation}

The calculated surface tension values are shown in Table I.
Notice that to relatively small changes of the scattering length $a_{12}$ correspond order of
magnitude changes of the surface tension $\sigma _{12}$. 
We will exploit such sensitivity to control the wetting properties of fluid (1,2),
as described in the following.
The surface tension $\sigma _{23}$ of the (2,3) surface is also shown,
which is much stronger than the reported values for the (1,2) system.

\begin{table}
\begin{tabular}{cccccccccc}
\hline
\hline
& $ a_{12}$$ (a_0) $  & $\sigma _{12}$$  (\mu K/\mu m^2) $ \\
\hline
 & $-40.5 $ & $8$  \\
 & $-42   $ & $26$  \\
 & $-44   $ & $80$  \\
 & $-46   $ & $187$  \\
 & $-48   $ & $346$  \\
 & $-50   $ & $563$  \\
 & $-52   $ & $825$  \\
 & $-54   $ & $1123$  \\
 & $-56   $ & $1452$  \\
\hline
\hline
& $ a_{23}$$ (a_0) $  & $\sigma _{23}$$  (\mu K/\mu m^2) $ \\
\hline
& $-200 $ & $1868$  \\
\hline
 \end{tabular}
  \caption{ Surface tension of the $(1,2)$ liquid as a function of the
scattering length $a_{12}$. The surface tension of the (2,3) liquid is also shown.
 \label{table1}
}
\end{table}

\begin{figure}[t]
\includegraphics[width=8cm]{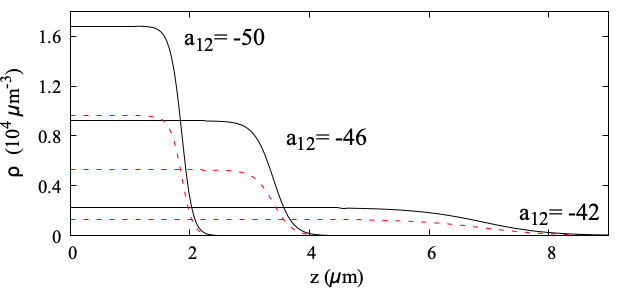}
\caption{Density profiles of the (1,2) slab for three different values of $a_{12}$ (for clarity, only one half  
is shown in the Figure. The solid (dashed) line shows the density of species 2 (1), respectively.
} 
\label{fig2}
\end{figure}

Fig.\ref{fig2} shows representative plots of the slab density profiles along the
direction perpendicular to the surface, for different values of $a_{12}$.
The two densities $\rho _1$ and $\rho _2$ (corresponding to the $^{23}$Na and $^{39}$K species,
respectively) are shown separately in the figure.
Notice the surface region becoming more diffuse as the scattering length approaches 
the threshold for droplet formation, with a corresponding lower surface tension.

We have checked, by doubling the width of the empty space separating the slab from its
repeated images, that the relative changes in the calculated values of the surface tensions 
$\sigma _{12}$ are negligible,
ranging from $10^{-3}$ when $a_{12}=-42\,a_0$, to $10^{-8}$ when $a_{12}=-44\,a_0$,
and to $10^{-12}$ when $a_{12}=-46\,a_0$.
The effect on the calculated value for the surface tension $\sigma _{23}$ is even smaller,
due to the narrower profile of the fluid-vacuum interface,
showing that interactions between periodic images have no tangible effect on the present
results. 


\subsection{Contact angle and wetting transition}

In order to study the wetting properties of the (1,2) fluid, we adopt 
a geometry resembling the usual set-up for wetting studies, i.e. 
a finite droplet of "liquid" adsorbed on a planar "solid" substrate. 
As stated previously, the role of the substrate is played here by the (2,3)
fluid, due to its greater stiffness, which is therefore in 
the form of a slab: the droplet will be adsorbed on the upper surface 
of the slab.

The (1,2) droplet used here has the form of a cylindrical cap,
i.e. a 2D geometry where the droplet is translationally 
invariant along one direction due to the use of periodic boundary conditions. 
The equilibrium configuration of the droplet-substrate system is found by propagating
Eq.\eqref{GP} in 
imaginary-time until convergence is achieved.

\begin{figure}[t]
\includegraphics[width=8cm]{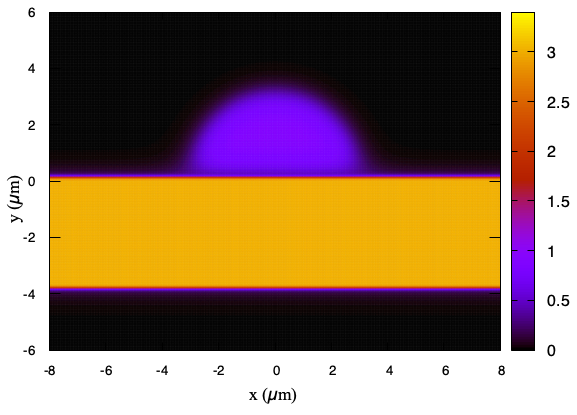}
\caption{ Density map in a plane perpendicular to the surface plane, 
showing the adsorbed (1,2) cylindrical droplet on top of a
slab (delimited by planar surfaces) made of the (2,3) fluid. Here $a_{12}=-44\,a_0$.
} 
\label{fig3}
\end{figure}

Fig.\ref{fig3} shows the equilibrium shape of a cylindrical (1,2) droplet
adsorbed on the (2,3) substrate, obtained with $a_{12}=-44\,a_0$.  
Notice that 
(i) the final cross-section of the adsorbed cap
is very close to a circular sector and
(ii) the substrate remains essentially unperturbed upon adsorption of the 
droplet, and maintains an almost undeformed planar shape.

Cylindrical droplets like those used here are often employed
in numerical simulations of wetting phenomena. Being
an essentially 2D simulation, the computational cost is largely reduced 
with respect to a fully 3D simulation of a finite spherical cap, therefore 
allowing for a deeper systematic study, and also
removing contact-line curvature dependencies which might be relevant for small droplets,
a cylindrical cap being the limit of a spherical one with infinitely large 
radius.

\begin{figure}[t]
\includegraphics[width=9cm]{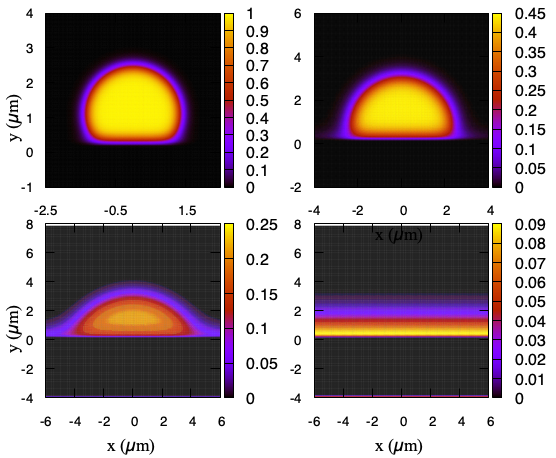}
\caption{Density maps of $\rho_1+\rho_2$ for the equilibrium configurations 
of cylindrical cap made of fluid (1,2) on
top of a planar slab of fluid (2,3), for different values of the 
scattering length $a_{12}$. From left to right, from top to bottom:
$a_{12}/a_0=-50,-45,-43,-42$.
} 
\label{fig4}
\end{figure}

Representative configurations are shown in Fig.\ref{fig4}, for different values of $a_{12}$.
For clarity, only the central portion of the system containing the (1,2) droplet is shown.
It appears that the contact angle diminishes as $a_{12}$ approaches 
the collapse value. At $a_{12}=-42\,a_0$ the fluid (1,2) has completely wetted the (2,3) substrate ($\theta =0$).

From these geometries, a reliable estimate for the contact angle $\theta $ can be obtained.
The latter is defined by $cos(\theta )=(y_3-y_0)/R$,
where $R$ is the radius of the cylindrical sector used to fit the
calculated densities and $y_0$ ($<0$) is the position of the center of the
circle: both $R$ and $y0$ are obtained from the fit. The position of the
contact plane $y_3$ ($>0$) is naturally defined here as the position 
of the Gibbs dividing surface of the density $\rho _3$, which represents the
substrate. In order to fit the 
circular profile of the (1,2) droplet, we make a least-squares fit to the 
points where the density component $\rho _1$ equals $\rho _1^c/2$, 
$\rho _1^c$ being the maximum density value of the droplet.

The extraction of the contact angle in the way described above 
involves a degree of arbitrariness, 
because both the contact plane and the liquid-vacuum contour 
have to be defined within diffuse liquid-vacuum interfaces. 
We have checked, however, that the resulting values of $\theta $ are not 
significantly affected by this choice. In particular, this check 
is important for the rather diffuse (1,2)-vacuum
interface (see Fig.\ref{fig2}). If instead of using the contour 
$\rho_1=\rho_1^{c}/2$ we use the value $\rho_1=\rho_1^{c}/4$, 
while keeping the contact plane fixed at the Gibbs dividing surface of the
sharper $\rho_3 $ interface, the 
calculated contact angle changes by $\sim 1-3\%$ (depending upon the value of 
$a_{12}$), showing  
that the location of the wetting transition is 
rather insensitive to the particular contour chosen for the fit.

\begin{figure}[t]
\includegraphics[width=8cm]{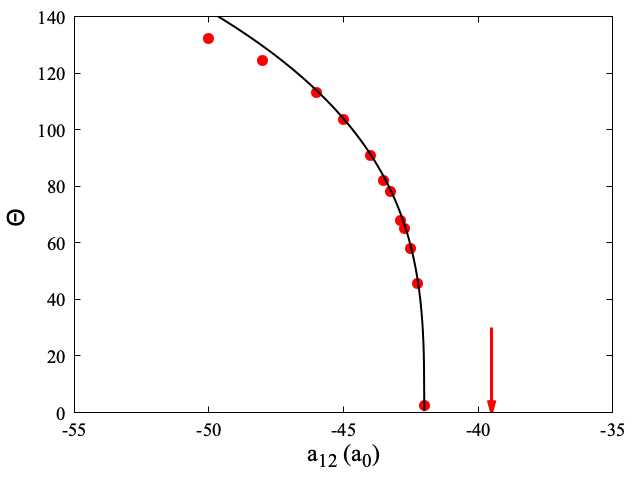}
\caption{Calculated values of the contact angles (in degrees) as a function of $a_{12}$.
The solid line shows the numerical fit described in the text.
The vertical arrow shows the threshold value below which self-bound states 
of the mixture (1,2) develop. 
} 
\label{fig5}
\end{figure}

The computed contact angle values as a function of $a_{12}$ are shown 
in Fig.\ref{fig5}, showing that a wetting transition occurs at $a_{12}=-42\,a_0$.

We notice that a similar behavior is found for the contact angle 
of a prototypical quantum fluid, liquid $^4$He, adsorbed on a solid surface:
calculations of the expected 
contact angle of a spherical $^4$He cap as a function
of the coupling constant describing the surface-liquid attraction indeed show for $\theta $ a
similar behavior \cite{anci98}.

The values of $\theta $ close to the wetting transition suggest a
power-law behavior in that region.
We thus made a fit of the contact angle values in the partial wetting region
(i.e. where $0<  \theta <90^\circ$) using the expression
$\theta = A(a_{12}^c-a_{12})^\gamma $, with the results
$A=72.12$, $a_{12}^c=-42.00$ and $\gamma =0.329$.
Notice that the exponent is very close to $1/3$.

To our knowledge, 
this cube-root behavior is not a standard prediction of the existing 
wetting theory for classical liquids, nor for Helium-4 or Bose mixtures. 
The published GP-based 
studies identify first-order wetting, critical wetting, 
and prewetting, but they do not report a universal $1/3$
exponent for the contact angle near the threshold. 


We do not interpret this exponent as universal; rather, it appears 
to reflect the microscopic dependence of the LHY-stabilized 
liquid-vacuum and liquid-liquid interfacial tensions on the scattering length.
Therefore, the observed cube-root law is most naturally 
interpreted as a non-universal signature of the microscopic 
quantum-liquid energetics of the two-component self-bound system.

\begin{figure}[t]
\includegraphics[width=8cm]{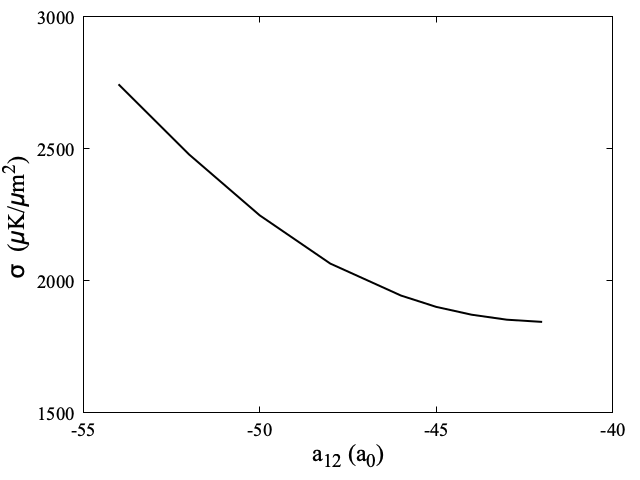}
\caption{Interface tension $\sigma _{SL}$, where S=(2,3) and L=(1,2),
as obtained using the calculated values for the contact angles in the Young's relation.
The units are the same of Table I.
} 
\label{fig6}
\end{figure}

We can use the computed contact angle values to extract the interfacial energy per
unit area of the (2,3) surface interacting with the (1,2) fluid, 
$\sigma _{SL} =\sigma _{SV} - \sigma _{LV} cos(\theta )$.
This is shown in Fig.\ref{fig6}.
A direct calculation of the interfacial tension $\sigma _{SL}$ is complicated by the fact that
in the partial wetting regime a planar contact between the two liquids is not stable,
the system forming instead a cap with finite contact angle. Thus the 
interfacial tension shown in Fig.\ref{fig6} should be regarded as the interfacial
energy (per unit surface) inferred from the Young construction.
Its smooth dependence on $a_{12}$ provides a consistency check of the capillary description, 
while a fully constrained planar-interface calculation is left for future work.

\subsection{Core-shell droplets in the complete-wetting regime}

\begin{figure}[t]
\includegraphics[width=8cm]{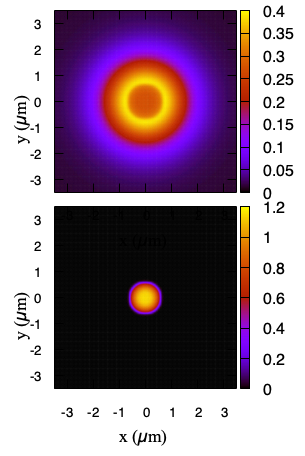}
\caption{
Column 
density, defined as the 3D atomic density \(\rho(x, y, z)\) 
integrated along the line of sight of a droplet with $a_{12}=-42\,a_0$.
The upper panel shows the projected density $\rho _1$, while the lower panel 
shows the projected density $\rho _3$. 
The projected density values are in units of $10^4 \mu m ^{-2}$.
Only a portion of the cell used in the actual simulation is shown for clarity.
} 
\label{fig7}
\end{figure}

From the findings reported in Fig.\ref{fig5} we now verify that, in the wetting regime
where $a_{12}\sim -42\,a_0$, a finite amount of (1,2) fluid interacting with 
a spherical drop made of (2,3) fluid will wrap around such a spherical core,
realizing a shell-shaped self-bound droplet made of (1,2) species.
One such equilibrium structure is shown in Fig.\ref{fig7}, where 
the 2D column-density image of species 1 is displayed (to simulate a species-selective
in-situ image of component 1 that could be performed in an experiment).
The displayed configuration has been obtained with a mixture of $N_1=6\times 10^4$, 
$N_2=8.7\times 10^4$, $N_3=1\times 10^4$ atoms, contained in a cubic simulation cell 
of sides $13\,\mu$m with a spatial mesh of $192^3$ points.
A hollow, shell-shaped structure would show, in such image, a ring-like projection 
with a central depletion, whereas 
a spherical droplet would appear as a filled disk with a centrally peaked density.
This is indeed what appears from the upper panel of Fig.\ref{fig7}, 
where the column-density image of species 1 displays the 
expected ring structure, whereas the 
lower panel, showing instead the column-density image of species 3, 
appears as a filled disk.

In order to avoid any bias during the imaginary-time propagation
towards the minimum energy configuration, we 
tried two different initial configurations:
(i) a uniform spherical mixture of the three species, and (ii) 
two adjacent (1,2) and (2,3) droplets (dimer-like configuration), and 
invariably ended with the same lowest-energy state, i.e. the one shown in Fig.\ref{fig7}.

\subsection{Vortical excitations in the $^{23}$Na-$^{39}$K hollow shell}

Due to its superfluid nature, the droplet of Fig.\ref{fig7} can host 
quantized vortical excitations when a sufficient 
amount of angular momentum is stored in it.
In particular,
the core-shell geometry prompts the question whether the outer
self-bound liquid can sustain quantized circulation. This question is
nontrivial because the coating phase is dilute and diffuse near the
wetting threshold, and because the total density is not hollow, the
central region being occupied by the denser $(2,3)$ liquid. 

The energy cost of a vortex line in a shell-shaped condensate
scales linearly with its thickness \cite{padavic}. Moreover, the lower the density of the
host medium, the lower the energy required to excite the vortex.
One therefore expects that the lowest-energy excitation possible in
the droplet shown in Fig.\ref{fig7} would be a 
linear, singly-quantized vortex in the species 1 only, with the other species not carrying 
any angular momentum.

We have searched for the lowest-energy vortex-like stationary state by
imprinting a singly quantized phase winding on different components of the
wave functions of the core-shell droplet, $ \psi_j({\bf r})\rightarrow \psi_j({\bf r})e^{i\phi}$,
where $\phi$ is the azimuthal angle around the vortex axis (which is taken here along $z$). 
The system is then propagated in imaginary time to find the minimum-energy configurations.
We have considered different possible vortical excitations:
a singly quantized vortex only in the species $j$, with $j=1,2,3$, and
also other (higher energy) configurations with vortices in two species
(1,2), (2,3) and (1,3).

We remark that 
these configurations should be distinguished from vortices in trapped
hollow condensates in 2D. In a 2D closed shell the topology favors
vortex-antivortex structures on the surface \cite{vort_2D}, whereas here the finite
thickness of the self-bound coating and the presence of a material core
allow singly quantized linear vortex excitations \cite{padavic}. 

A useful quantitative characterization is provided by the vortex
excitation energy
\begin{equation}
\Delta E_v = E_v-E_0,
\end{equation}
where $E_v$ and $E_0$ are the total energies of the configuration with and
without vortices.

These quantities determine the approximate rotation frequency
$\Omega_c=\Delta E_v/L_z$ above which the vortex becomes energetically
favored in a rotating frame. 
For the above configurations with one vortex line in the 
$j$-species $L_z=N_j\hbar $, while $L_z=(N_i+N_j)\hbar $
for the configurations hosting a vortex line in two species $i,j$.

We report in Table II the calculated frequencies for different vortex configurations
($v_{i}$ meaning a single vortex line in the species $i$, while $v_{ij}$ means
a vortex line in the species $i$ and one in the species $j$).
Notice that configurations $v_{2} $ and $v_{12}$ turn out to be unstable, 
i.e. during imaginary-time relaxation 
the imprinted phase singularity is expelled from the droplet, which
eventually relaxes to the vortex-free configuration with zero angular momentum
shown in Fig.\ref{fig7}.

As expected, the lowest-energy excitation in Table II 
corresponds to a singly quantized vortex line 
imprinted in the species $1$ only.

\begin{table}
\begin{tabular}{cccccccccc}
\hline
\hline
&   & $L_z/\hbar$  & $\omega /2\pi $ $ (Hz)$ \\
\hline
 & $v_{1} $ &$N_1$ & $179$  \\
 & $v_{2}  $ & $N_2$&$unstable$  \\
 & $v_{3}  $ & $N_3$&$2540$  \\
 & $v_{12}  $ & $N_1+N_2$&$unstable$  \\
 & $v_{13}  $ & $N_1+N_3$&$516$  \\
 & $v_{23}  $ & $N_2+N_3$&$738$  \\
\hline
\hline
 \end{tabular}
  \caption{ Calculated frequencies for vortex nucleation in the droplet of Fig.\ref{fig7}.
 \label{table2}
}
\end{table}

The
equilibrium density pattern for the lowest-energy configuration $v_{1}$ is 
shown in Fig.~8. The density depletion in
$\rho_1$ identifies a singly quantized vortex core localized in the
outer shell. The line avoids penetrating the dense $(2,3)$ core (not shown), so that
the vortex is predominantly an excitation of the coating liquid. The
corresponding column density, Fig.~9, shows how the vortex would appear
experimentally in species-selected projection of the density $\rho_1$ along the line of sight.

The vortex calculations reported above should be regarded as illustrative. 
They show that the component-resolved shell geometry can accommodate 
quantized circulation in the outer component within the present mean-field-plus-LHY model. 
A full analysis of dynamical stability, nucleation pathways, vortex bending, 
and lifetimes is beyond the scope of this work.

\begin{figure}[t]
\includegraphics[width=8cm]{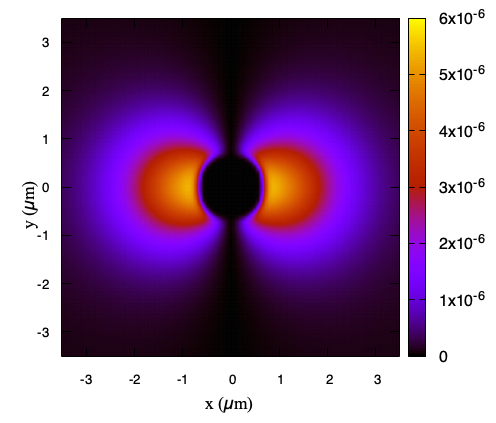}
\caption{Density map $\rho _1$ in the central plane of the shell
hosting a vortex line.
The density values are in units of $10^4 \mu m ^{-3}$.
Only a portion of the cell used in the actual simulation is shown for clarity.
} 
\label{fig8}
\end{figure}

\begin{figure}[t]
\includegraphics[width=8cm]{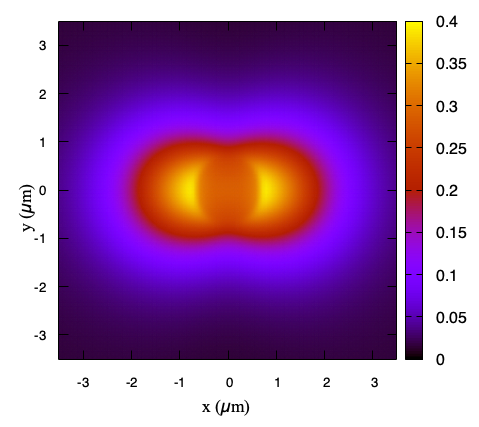}
\caption{ Projected density $\rho _1$ of the shell along the line of sight
for the configuration shown in Fig.\ref{fig8}.
The projected density values are in units of $10^4 \mu m ^{-2}$.
Only a portion of the cell used in the actual simulation is shown for clarity.
} 
\label{fig9}
\end{figure}

\section{Conclusions}

We have studied the wetting properties of two self-bound quantum liquids 
realized within a three-component $^{23}$Na-$^{39}$K-$^{41}$K Bose mixture. 
The two relevant fluids are the binary droplets formed by components $(1,2)$ and $(2,3)$. 
For the parameters considered here, the $(2,3)$ liquid has a much larger surface 
tension and behaves as an effectively rigid substrate for the softer $(1,2)$ liquid. 
This separation of stiffness makes it possible to formulate the problem 
in close analogy with conventional capillarity, with vacuum playing the role of the vapor phase.

By varying the scattering length $a_{12}$, we have shown that the surface 
tension of the $(1,2)$ liquid changes by orders of magnitude with relatively small
changes of $a_{12}$. This provides a
sensitive control knob for the wettability of the $(2,3)$ substrate. 
Calculations for cylindrical-cap droplets adsorbed on a planar $(2,3)$ 
slab show that the contact angle decreases as $a_{12}$ approaches the 
threshold for self-binding and vanishes at approximately $a_{12}^{c}\simeq -42\,a_0$, 
signaling a transition to complete wetting. The calculated contact angles are well 
described by an empirical power law with an effective exponent 
close to $1/3$. Although this exponent is not expected to be universal, 
it suggests a nonanalytic dependence of the interfacial free-energy 
balance on the microscopic interaction strengths.

The complete-wetting regime has an immediate consequence for 
droplets. A finite amount of the $(1,2)$ liquid can wrap a 
spherical $(2,3)$ core, producing a shell-shaped self-bound droplet in free space. 
The resulting structure is qualitatively different from shell-shaped 
condensates produced by external trapping potentials: here the 
shell is stabilized by the balance of mean-field attraction, 
quantum-fluctuation pressure, and interfacial energies. 
A clear experimental signature could be revealed by 
measuring the column density of species 1,
which should display a telltale ring-like projection
with the core component remaining centrally peaked.
We have shown that the shell can host a singly quantized vortex, 
illustrating the possibility of studying superfluid circulation 
in a curved, multiply connected quantum liquid.


The present results indicate that wetting phenomena in 
quantum liquids offer a plausible and controllable route
to designing multicomponent self-bound quantum liquids 
with controlled geometry and topology.


Some comments are in order at this point.
Strictly speaking, the self-bound spherical structure obtained in the complete-wetting 
regime is not hollow in its total density, since the central region is 
occupied by the (2,3) liquid and the shared component 2 is present 
both in the core and in the shell. The term “shell-shaped” should therefore 
be understood in a component-resolved sense: component 1, and the (1,2) liquid 
associated with it, form an outer layer surrounding the (2,3) core. 
A species-selective image of component 1 would display a ring-like 
column density with a central depletion, whereas the total density 
remains filled. In this sense the system realizes a self-bound core-shell 
quantum droplet rather than an empty-cavity hollow droplet. 
This distinction is important for possible applications to 
curved or multiply connected quantum fluids, since the relevant topology 
depends on which component, or which superfluid mode, is being probed.

A natural direction for future work is to search for more favorable 
atomic mixtures and interaction regimes in which complete wetting 
occurs farther from the self-binding threshold of the adsorbed liquid. In the present 
realization, 
the wetting transition is reached only when the scattering length $a_{12}$
is tuned close to the threshold for formation of the (1,2) droplet. 
As a consequence, the wetting liquid has a low equilibrium density, 
a small surface tension, and a relatively large healing length, 
producing a diffuse shell rather than a sharp, thin quantum film. 
While this is sufficient to demonstrate the wetting mechanism and 
the formation of a shell-shaped self-bound state, it is not the most 
favorable limit for studying curved-space or topological superfluid effects, 
where a thinner and denser shell would be preferable. 
Future studies should therefore explore other three-component mixtures, 
different choices of the two binary liquids, and alternative Feshbach-tuned interaction paths, 
with the aim of independently controlling the wettability and the intrinsic 
density/stiffness of the outer liquid. An ideal system would 
combine complete wetting with a sizable equilibrium density and 
surface tension of the coating phase, allowing the formation of 
thinner, more mechanically robust shells with better separated radial 
and angular length scales. Such an optimization would turn the 
present proof-of-principle mechanism into a more practical 
platform for realizing shell-shaped quantum liquids with 
controlled curvature and topology.

\bigskip

\acknowledgments
The author wishes to thank A. Burchianti, C. Fort and G. Mistura for useful comments.
This work is supported by the University of Padova under the BIRD 2025 project 
"Interacting Quantum Mixtures: From Droplets to Supersolids".


\begin{thebibliography}{99}


\bibitem{petrov_15} D. S. Petrov, 
{\it Quantum Mechanical Stabilization of a Collapsing Bose-Bose Mixture},
Phys. Rev. Lett. {\bf 115}, 155302 (2015).

\bibitem{kadau} H. Kadau, M. Schmitt, M. Wenzel, C. Wink, T. Maier, I. Ferrier-Barbut and T. Pfau,
{\it Observing the Rosensweig instability of a quantum ferrofluid},
Nature {\bf 530}, 194 (2016).

\bibitem{ferrier} I. Ferrier-Barbut, H. Kadau, M. Schmitt, M. Wenzel and T. Pfau,
{\it Observation of Quantum Droplets in a Strongly Dipolar Bose Gas},
Phys. Rev. Lett. {\bf 116}, 215301 (2016).

\bibitem{Schmitt2016} 
M. Schmitt, M. Wenzel, F. B{\"o}ttcher, I. Ferrier-Barbut and T. Pfau, 
{\it Self-bound droplets of a dilute magnetic quantum liquid},
Nature {\bf 539}, 259-262 (2016).

\bibitem{Ferlaino2016}
L. Chomaz, S. Baier, D. Petter, M.J. Mark, F. W\"achtler, L. Santos and F. Ferlaino, 
{\it Quantum-Fluctuation-Driven Crossover from a Dilute Bose-Einstein 
Condensate to a Macrodroplet in a Dipolar Quantum Fluid},
Phys. Rev. X, {\bf 6}, 041039 (2016).

\bibitem{Cabrera2018}
C.R. Cabrera, L. Tanzi, J. Sanz, B. Naylor, P. Thomas, P. Cheiney and L. Tarruell,
{\it Quantum liquid droplets in a mixture of Bose-Einstein condensates},
Science, {\bf 359}, 301-304 (2018).

\bibitem{Semeghini2018}
G. Semeghini, G. Ferioli, L. Masi, C. Mazzinghi, L. Wolswijk, F. Minardi, 
M. Modugno, G. Modugno, M. Inguscio and M. Fattori, 
{\it Self-Bound Quantum Droplets of Atomic Mixtures in Free Space},
Phys. Rev. Lett. {\bf 120}, 235301 (2018).

\bibitem{Derrico2019} C. D’Errico, A. Burchianti, M. Prevedelli, L. Salasnich,
F. Ancilotto, M. Modugno, F. Minardi, and C. Fort, 
{\it Observation of quantum droplets in a heteronuclear bosonic
mixture}, Phys. Rev. Res. {\bf 1}, 033155 (2019).

\bibitem{Guo2021}
Z. Guo, F. Jia, L. Li, Y. Ma, J.M. Hutson, X. Cui and D. Wang,
{\it Lee-Huang-Yang effects in the ultracold mixture of $^{23}\mathrm{Na}$ 
and $^{87}\mathrm{Rb}$ with attractive interspecies interactions},
Phys. Rev. Res. {\bf 3}, 033247 (2021).

\bibitem{review} F. Bottcher, J.-N. Schmidt, J. Hertkorn, K. S. H. Ng, S. D. Graham, 
M. Guo, T. Langen and T. Pfau, 
{\it New states of matter with fine-tuned interactions: quantum droplets and dipolar supersolids}, 
Rep. Prog. Phys. {\bf 84}, 012403 (2021).

\bibitem{lhy}T.D. Lee, K. Huang, and C.N. Yang, 
{\it Eigenvalues and Eigenfunctions of a Bose System of Hard Spheres and Its Low-Temperature Properties},
Phys. Rev. {\bf 106}, 1135 (1957).

\bibitem{anci_mult}
L. Cavicchioli, C. Fort, F. Ancilotto, M. Modugno, F. Minardi 
and A. Burchianti, 
{\it Dynamical Formation of Multiple Quantum Droplets in a Bose-Bose Mixture}, 
Phys. Rev. Lett. {\bf 134}, 093401 (2025).

\bibitem{anci23} F. Ancilotto, M. Barranco and M. Pi,
{\it Breakup of quantum liquid filaments into droplets},
Phys. Rev. A {\bf 107}, 063312 (2023);
F. Ancilotto, M. Modugno and C. Fort,
{\it Suppression of capillary instability in a confined quantum liquid filament},
Phys. Rev. A {\bf 112}, 043316 (2025).

\bibitem{borromean}
Y. Ma, C. Peng and X. Cui,
{\it Borromean Droplet in Three-Component Ultracold Bose Gases},
Phys. Rev. Lett. {\bf 127}, 043002 (2021).

\bibitem{ma}Y. Ma and X. Cui,
{\it Shell‑shaped quantum droplet in a three‑component ultracold Bose gas},
Phys. Rev. Lett. {\bf 134}, 043402 (2025).

\bibitem{comment} F. Ancilotto, 
{\it Comment on “Shell-Shaped Quantum Droplet in a Three-Component Ultracold Bose Gas”},
Phys. Rev. Lett. {\bf 135}, 159301 (2025). 

\bibitem{anci25} F. Ancilotto,
{\it Quantum diatomic chain: A supersolid structure in a three-component Bose mixture},
Phys. Rev. A {\bf 112}, 063317 (2025).

\bibitem{carollo1} R. A. Carollo, D. C. Aveline, B. Rhyno, S. Vishveshwara, 
C. Lannert, J. D. Murphree, E. R. Elliott, J. R.
Williams, R. J. Thompson and N. Lundblad, 
Nature {\bf 606}, 281 (2022).

\bibitem{jia} F. Jia, Z. Huang, L. Qiu, R. Zhou, Y. Yan, and D. Wang,
{\it Expansion Dynamics of a Shell-Shaped Bose-Einstein Condensate},
Phys. Rev. Lett. {\bf 129}, 243402 (2022).

\bibitem{huang1} Z. Huang, K. Y. Lee, C. K. Wong, L. Qiu, B. Yang,
Y. Yan and D. Wang,
{\it Probing the hollowing transition of a shell-shaped Bose-Einstein condensate with collective excitation},
Phys. Rev. Res. {\bf 7}, 033056 (2025).


\bibitem{rhyno} 
B. Rhyno, K. Sun, J. Bedessem, N. Gaaloul, N. Lundblad and S. Vishveshwara,
{\it  Shell-shaped Bose-Einstein condensates: Dynamics, excitations, and thermodynamics },
arXiv:2512.13858 (2026).

\bibitem{sun} K. Sun, K. Padavic, F. Yang, S. Vishveshwara and C. Lannert, 
{\it Static and dynamic properties of shell-shaped condensates}, 
Phys. Rev. A {\bf 98}, 013609 (2018); 
C. Lannert, T.-C. Wei and S. Vishveshwara, 
{\it Dynamics of condensate shells: Collective modes and expansion}, 
Phys. Rev. A {\bf 75}, 013611 (2007).

\bibitem{salasnich} A. Tononi and L. Salasnich, 
{\it Low-dimensional quantum gases in curved geometries}, 
Nat. Rev. Phys. {\bf 5}, 398 (2023).

\bibitem{tononi} A. Tononi and L. Salasnich,
{\it Shell-shaped atomic gases},
Physics Reports Volume {\bf 1072}, 1-48 (2024).

\bibitem{Ral60}
A. Ralston and H. S. Wilf, {\it Mathematical methods for digital computers} (John Wiley and Sons, New York, 1960).

\bibitem{bonn09} D. Bonn, J. Eggers, J. Indekeu, J. Meunier and E. Rolley,
{\it Wetting and spreading},
Rev. Mod. Phys. {\bf 81}, 739 (2009).

\bibitem{wet-he_exp}
P. J. Nacher and J. Dupont-Roc, 
{\it Experimental evidence for nonwetting with superfluid helium}, 
Phys. Rev. Lett. {\bf 67}, 2966–2969 (1991);
K. S. Ketola, S. Wang and R. B. Hallock, 
{\it Anomalous wetting of helium on cesium},
Phys. Rev. Lett. {\bf 68}, 201–204 (1992);
G. Mistura, Hyun C. Lee, M.H.W. Chan,
{\it Quartz microbalance study of hydrogen and helium adsorbed on
a Rubidium surface},
Physica B: Condensed Matter {\bf 194-196}, 661 (1994);
J. E. Rutledge and P. Taborek, 
{\it Prewetting phase diagram of $^4$He on cesium},
Phys. Rev. Lett. {\bf 69}, 937–940 (1992);
D. Reinelt, J. Klier and P. Leiderer, 
{\it Wetting studies of liquid $^4$He on various Cs surfaces},
J. Low Temp. Phys. {\bf 113}, 805–810 (1998);
M. Barranco, M. Guilleumas, E.S. Hernandez, R. Mayol, M. Pi and L. Szybisz,
{\it From nonwetting to prewetting: The asymptotic behavior of $^4$He drops on alkali substrates},
Phys. Rev. B {\bf 68}, 024515 (2003).

\bibitem{wet-he_th}
E. Cheng, M. W. Cole, W. F. Saam and J. Treiner, 
{\it Helium prewetting and nonwetting on weak-binding substrates},
Phys. Rev. Lett. {\bf 67}, 1007–1010 (1991);
F. Ancilotto, F. Faccin and F. Toigo, 
{\it Wetting transitions of $^4$He on alkali-metal surfaces from density-functional calculations},
Phys. Rev. B {\bf 62}, 17035 (2000).

\bibitem{ind04} J. O. Indekeu and B. Van Schaeybroeck, 
“Extraordinary wetting phase diagram for mixtures of Bose–Einstein condensates”, 
Phys. Rev. Lett. {\bf 93}, 210402 (2004).

\bibitem{sch15} B. Van Schaeybroeck and J. O. Indekeu, 
“Critical wetting, first-order wetting and prewetting phase 
transitions in binary mixtures of Bose–Einstein condensates”, 
Phys. Rev. A {\bf 91}, 013626 (2015).

\bibitem{sch22} B. Van Schaeybroeck, P. Navez, and J. O. Indekeu, 
“Interface potential and line tension for Bose–Einstein condensate 
mixtures near a hard wall”, 
Phys. Rev. A {\bf 105}, 053309 (2022).

\bibitem{song} N.V. Thu, T.H. Phat and P.T. Song,
{\it Wetting phase transition of two segregated 
Bose–Einstein condensates restricted by a hard wall},
Physics Letters A {\bf 380}, 1487 (2016).

\bibitem{anci98} F. Ancilotto, A.M. Sartori and F. Toigo, 
{\it Structure and contact angle of liquid $^4$He 
droplets on a Cs surface},
Phys. Rev. B {\bf 58}, 5085 (1998).

\bibitem{padavic} K. Padavic, K. Sun, C. Lannert and Smitha Vishveshwara,
{\it Vortex-antivortex physics in shell-shaped Bose-Einstein condensates},
Phys. Rev. A {\bf 102}, 043305 (2020).

\bibitem{vort_2D} A.M. Turner, V. Vitelli and D.R. Nelson, 
{\it Vortices on curved surfaces}, 
Rev. Mod. Phys. {\bf 82}, 1301 (2010);
S.J. Bereta, M.A. Caracanhas and A.L. Fetter, 
{\it Superfluid vortex dynamics on a spherical film}, 
Phys. Rev. A {\bf 103}, 053306 (2021).

\end{thebibliography}
\end{document}